\documentclass[]{aa}
\usepackage{txfonts}
\usepackage{graphicx}
\usepackage{natbib}
\usepackage{color}

\usepackage[hidelinks=true]{hyperref}
\definecolor{dark-red}{rgb}{0.4,0.15,0.15}
\definecolor{dark-blue}{rgb}{0.1,0.1,0.6}
\definecolor{medium-blue}{rgb}{0,0,0.5}
\hypersetup{
    colorlinks, linkcolor={dark-red},
    citecolor={dark-blue}, urlcolor={medium-blue}
}

\graphicspath{{/net/reko/work1/soe/rohan/SP-hinode/ca_jets/figures/}{/net/reko/work1/soe/rohan/SP-hinode/ca_jets/figures/}}

\begin{document}

\title{Small-scale chromospheric jets above a sunspot light bridge}

\author{Rohan E. Louis$^{1}$ \and Christian Beck$^{2}$ \and
Kiyoshi Ichimoto$^{3}$}


\institute{Leibniz-Institut f\"ur Astrophysik Potsdam (AIP),
	  An der Sternwarte 16, 14482 Potsdam, Germany \and
          National Solar Observatory, Sacramento Peak, 
          3010 Coronal Loop, Sunspot, New Mexico 88349, U.S.A. \and
          Kwasan and Hida Observatories, Kyoto University, Yamashina, 
          Kyoto 607-8471, Japan}

\date{Received ... 2012 / Accepted ...}

\abstract
   {The chromosphere above sunspot umbrae and penumbrae shows several different
types of fast dynamic events such as running penumbral waves, umbral flashes, and penumbral microjets.}
   {The aim of this paper is to identify the physical driver responsible for the dynamic and small-scale chromospheric jets 
above a sunspot light bridge.}
   {High-resolution broadband filtergrams of active region NOAA 11271 in Ca~{\sc ii}~H and G~band were obtained with the 
Solar Optical Telescope on board {\em Hinode}. We identified the jets in the Ca~{\sc ii}~H images using a semi-automatic 
routine and determined their length and orientation. We applied local correlation tracking (LCT) to the G-band images to 
obtain the photospheric horizontal velocity field. The magnetic field topology was derived from a Milne-Eddington 
inversion of a simultaneous scan with the Spectropolarimeter.}
   {The chromospheric jets consist of a bright, triangular-shaped blob that lies on the light bridge, while the apex 
of this blob extends into a spike-like structure that is bright against the dark umbral background. Most of the 
jets have apparent lengths of less than 1000 km and about 30\% of the jets have lengths between 1000--1600 km. 
The jets are oriented within $\pm$35$^\circ$ to the normal of the spine of the light bridge. Most 
of them are clustered near the central part of the light bridge within a 2\arcsec area. The jets are seen to move 
rapidly along the light bridge and many of them cannot be identified in successive images
taken with a 2 min cadence. The jets are primarily located on one side of the light bridge and are directed into the 
umbral core. The Stokes profiles at or close to the location of the blobs on the LB exhibit both a significant net 
circular polarization and 
multiple components, including opposite-polarity lobes. The magnetic field diverges from the light bridge towards the 
umbral cores that it separates. The LCT reveals that in the photosphere there is a predominantly uni-directional flow 
with speeds of 100--150 m~s$^{-1}$ along the light bridge. This unidirectional flow is interrupted by a patch of weak 
or very small motions on the light bridge which also moves along the light bridge.} 
    {The dynamic short-lived chromospheric jets above the LB seem to be guided by the magnetic 
field lines. Reconnection events are a likely trigger for such phenomenon since they occur at locations where the
magnetic field changes orientation sharply and where we also observe isolated patches of opposite-polarity magnetic 
components. We find no clear relation between the jets and the photospheric flow pattern.}


\maketitle

\section{Introduction}
\label{intro}
Light bridges (LBs) are bright structures that intrude and/or divide umbrae of sunspots and pores into multiple 
smaller umbral cores. Light bridges can have a penumbral or granular morphology \citep{1979SoPh...61..297M} and typically 
show a dark lane running along their central axis in high-resolution observations
\citep{1994ApJ...426..404S,2002A&A...383..275H,2003ApJ...589L.117B,2004SoPh..221...65L,2008ApJ...672..684R}. At 
the photosphere, LBs can be perceived as field-free intrusions of hot plasma into the gappy umbral magnetic 
field \citep{1979ApJ...234..333P,1986ApJ...302..809C} or large-scale magneto-convective structures 
\citep{2008ApJ...672..684R}. In either of the above cases, the LB forces the neighbouring umbral magnetic field 
to form a canopy around it which leads to a stressed magnetic configuration \citep{2006A&A...453.1079J}, particularly 
in the chromosphere and transition region. Such a topology has been suggested to be responsible for a wide variety 
of chromospheric activity such as surges in H$\alpha$ \citep{2001ApJ...555L..65A}, strong brightenings and ejections 
observed in Ca~{\sc ii}~H \citep{2008SoPh..252...43L,2009ApJ...696L..66S,2009ApJ...704L..29L,2011ApJ...738...83S}, or 
brightness enhancements in the transition region \citep{2003ApJ...589L.117B}. 

In this paper, we analyse the properties of small-scale chromospheric jets in  a sunspot LB, using broadband 
filtergrams and spectropolarimetric data from {\em Hinode} \citep{2007SoPh..243....3K}. We relate 
the properties of the jets to the photospheric flow field and the magnetic field topology to identify what drives 
them and governs their evolution.

\section{Observations}
\label{data}
For this investigation, we utilised broadband filtergrams of the active region NOAA 11271 acquired by the Solar 
Optical Telescope \citep[SOT,][]{2008SoPh..249..167T} on board {\em Hinode}.  On 2011 August 19, the SOT observed 
the active region in the G band (using a 0.8 nm filter centred at 430.5 nm) and Ca~{\sc ii}~H (using a 0.3 nm 
filter centred at 396.9 nm) with a spatial sampling of 0\farcs11 and a cadence of 2 min for a duration of nearly 
2.5 hr from 8:05--10:33~UT. The field of view (FOV) of the filtergrams was 111\arcsec$\times$111\arcsec. The active 
region was located at a heliocentric angle $\Theta$ of 29$^\circ$ at the time of the observations. The Ca~{\sc ii}~H  
images were recorded after the G-band images with a delay of 4 s. All images were reduced with the standard routines 
of the Solar-Soft package. The images were subsequently co-aligned using a two-dimensional cross-correlation routine.  

At the same time, the FOV was scanned with the Hinode spectropolarimeter 
\citep[SP;][]{2008SoPh..249..233I,2013SoPh..283..579L}. 
During the time of the filtergram acquisition, the SP took five scans of the active 
region in the fast mode. Each scan covered a FOV of 75\arcsec$\times$82\arcsec. In the fast mode, the SP 
recorded the four Stokes profiles of the Fe {\sc i} lines at 630 nm with a spectral sampling of 21.5~m\AA, a step 
width of 0\farcs29, and a spatial sampling of 0\farcs32 along the slit. Routines included in the 
Solar-Soft package were employed to reduce the Level 0 data \citep{2013SoPh..283..601L}.
We used the corresponding Level 2 maps of the vector magnetic field from MERLIN\footnote{Level 2 maps from MERLIN 
inversions are provided by the Community Spectro-polarimetric Analysis Center at the  
link http://www.csac.hao.ucar.edu/csac} \citep{2007MmSAI..78..148L} inversions of the SP data. 
MERLIN uses a Milne-Eddington atmosphere where all physical parameters are constant with height 
except for the source function that varies linearly with optical depth. MERLIN assumes a single magnetic component 
and a variable stray light factor. The following parameters are retrieved by the code: field strength, 
inclination, azimuth, line-of-sight (LOS) velocity, damping constant, doppler width, line strength, macroturbulence, 
source function, source function gradient, and a stray light factor. 
The LOS observables were transformed to the local-reference-frame. Prior to the transformation, the 180$^\circ$ 
ambiguity in the azimuth was resolved manually by assuming a radial orientation of the transverse magnetic field and a 
spatially smooth azimuth.

\begin{figure}[!h]
\centerline{
\includegraphics[angle=0,width = 0.98\columnwidth]{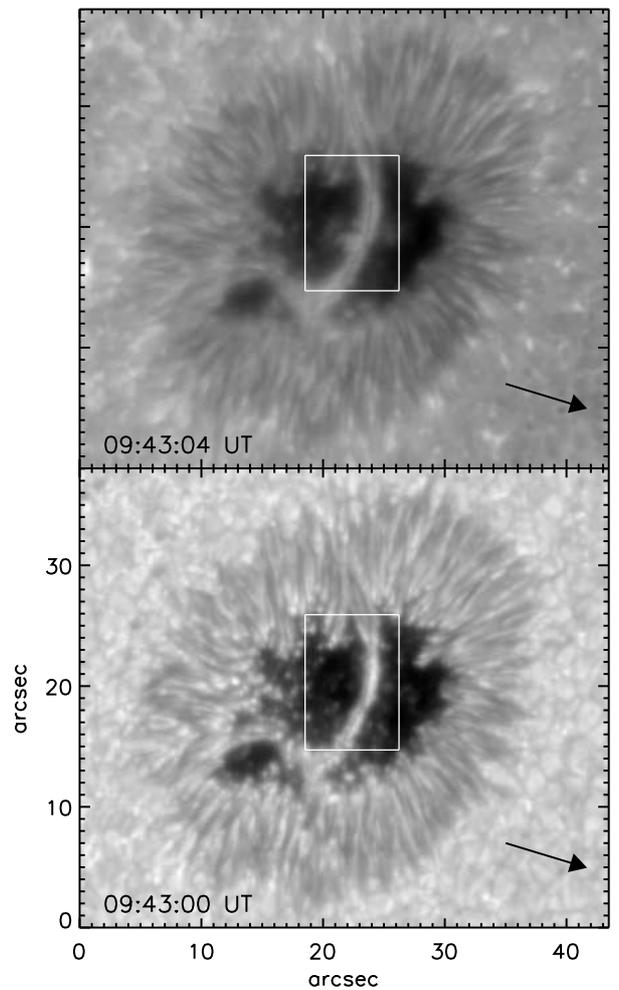}
}
\vspace{-5pt}
\caption{Leading sunspot in NOAA AR 11271. The top and bottom panels correspond to the Ca~{\sc ii}~H and G band, 
respectively. The large white rectangle represents the FOV selected for the analysis and shown in subsequent 
figures. The arrow points to disk centre. The images are displayed on a logarithmic scale.}
\label{fig01}
\end{figure}

\section{Data analysis}
\subsection{Local correlation tracking}
\label{track}
To determine if the proper motions have a role in the excitation of the chromospheric jets, we 
analysed the horizontal flow in the photosphere derived from local correlation tracking 
\citep{1986ApOpt..25..392N,1988ApJ...333..427N,2004ApJ...610.1148W,2008ASPC..383..373F}. 
Local correlation tracking (LCT) computes 
the relative displacement of small subregions centred on a particular pixel with subpixel accuracy using 
cross-correlation techniques. A Gaussian window, whose full width at half maximum needs to be roughly the 
size of the structures that are to be tracked, is used to apodize the subregions. Thus, the horizontal 
speed at each pixel can be determined knowing the displacement and the time interval. First, the G-band 
time sequence was filtered for acoustic waves using a phase velocity cutoff value of 7 km~s$^{-1}$ 
\citep{1989ApJ...336..475T}. After several trials, an apodizing window with a width of 1\arcsec~and 
a time difference of 2 min were chosen. Twenty successive velocity images were averaged to reduce the noise 
in the measurements. The LCT parameters are similar to the ones used in \citet{2010A&A...516A..91V} and 
\citet{2012ApJ...755...16L}.

\subsection{Identification of chromospheric jets}
\label{ident}
\begin{figure*}[!ht]
\centerline{
\includegraphics[angle=90,width = 1.\textwidth]{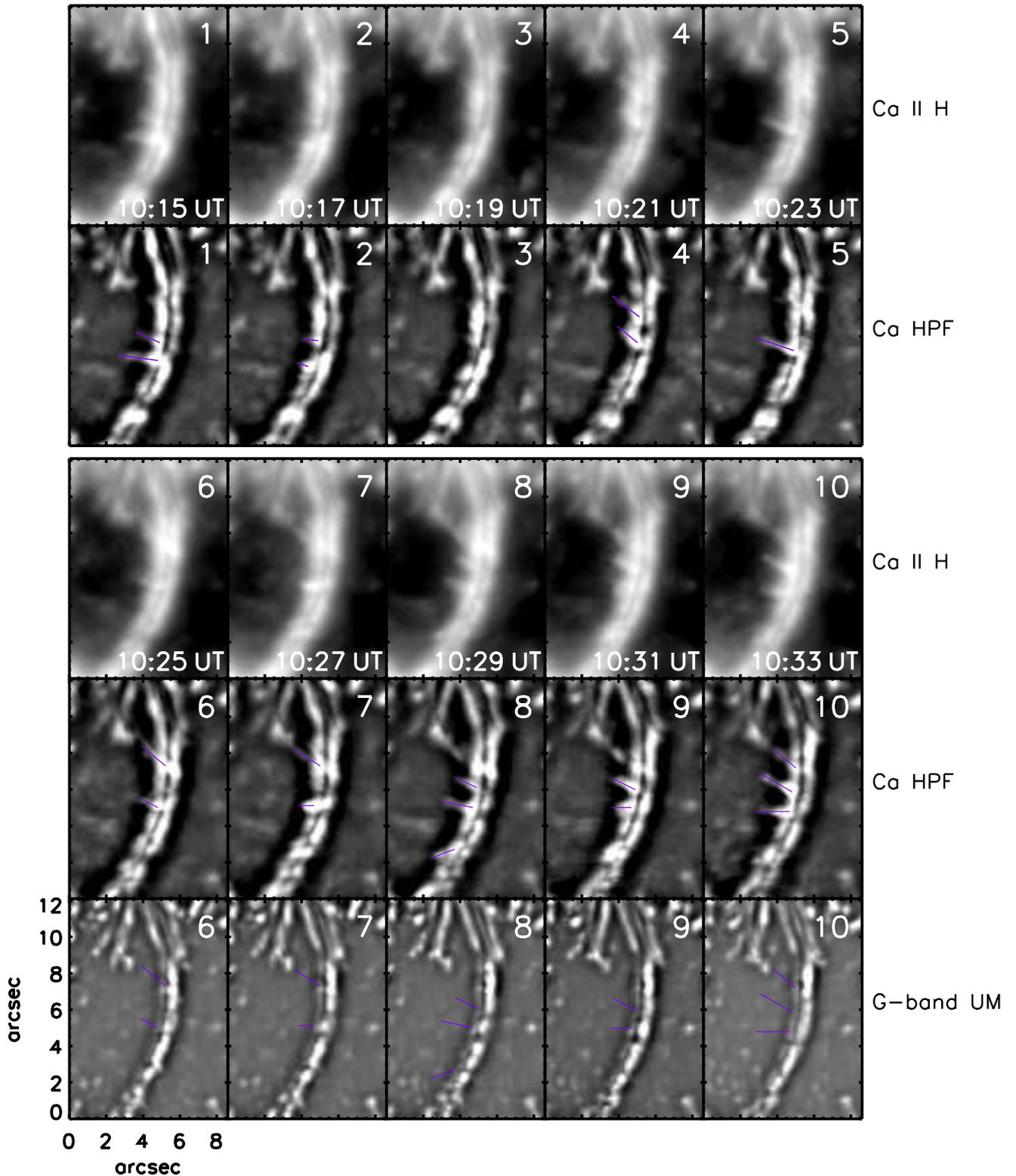}
}
\vspace{-45pt}
\caption{Chromospheric jets seen in Ca~{\sc ii}~H. The top panels correspond to the Ca~{\sc ii}~H time 
sequence starting at 10:15 UT. The bottom panels are the corresponding high-pass filtered images (Ca HPF), with 
the purple lines representing the jets identified using the semi-automatic algorithm. The bottom panel shows 
unsharp masked G-band (G-band UM) images that correspond to the Ca~{\sc ii}~H time sequence shown in the panel 
above. }
\label{fig02}
\end{figure*}

The top and bottom panels of Fig.~\ref{fig01} show the leading sunspot in active region (AR) NOAA 11271 in 
Ca~{\sc ii}~H and in the G band, respectively. An animation of the Ca~{\sc ii}~H and G-band images of the full 
observation is available in the online section. The larger LB enclosed in the white rectangle was selected 
for the analysis. We chose the above active region because the LB was stable for a duration of more than 48 hrs 
and was located inside a regular sunspot. Because of the reduced X-band telemetry of {\em Hinode} after the end of 
2007, the cadence of the imaging data was limited to allow for a reduction in data volume. Quantities 
such as lifetimes, proper motion and speeds could not be determined for most of the small-scale jets 
because they do not appear in consecutive frames (see Fig~\ref{fig02}). The animation of the 
time series shows a few longer-lived jets, as well as some cases where jets seem to move rapidly along the axis of 
the LB nearly in the same direction as the horizontal flow observed in the photosphere. Following the 
limitation caused by the temporal sampling, we focused on the characteristic properties of 
the jets, namely, position, length, and orientation as a statistical sample at an arbitrary time during their 
evolution.

The chromospheric jets on the LB are seen as a combination of bright, triangular blobs on the LB itself 
and spike-like structures which extend outward into the adjacent umbral background (Fig.~\ref{fig02}). We 
refer to the blobs on the LB as the base of the jets. Identifying the jet required locating the base, 
followed by determining the orientation and length of the jet. The above steps were performed in a 
semi-automatic manner. Each Ca~{\sc ii}~H image was high-pass filtered using an 11-pixel boxcar as 
shown in the bottom panels of Fig.~\ref{fig02}. The base of the jet located on the LB was selected by 
eye. A 7$\times$7 pixel window centred on the selected position was used to determine the position of 
maximum intensity. Once the base of the jet was identified, the rest of the procedure was performed by 
the algorithm described below. 

The orientation of the jet was determined by the angle at which the mean intensity along a specific radial 
distance from the base of the jet had a maximum value. To do this, we employed the loop-directivity procedure of 
\citet{2010SoPh..262..399A} that is part of a routine used for identifying coronal loops. The mean intensity 
along a given azimuthal direction ($\phi_k$) for a fixed radial length ($L$) is given by 
\begin{eqnarray}
I_k &=& \frac{1}{L} \sum_{l=0}^{L} I(x_{kl},y_{kl}) \textrm{ , where} \\
x_{kl} &=& x_b + l\cos{\phi_k} \textrm{ , and} \\
y_{kl} &=& y_b + l\sin{\phi_k}{\,.}
\end{eqnarray}

\begin{figure}[!h]
\centerline{
\hspace{10pt}
\includegraphics[angle=0,width = 0.8\columnwidth]{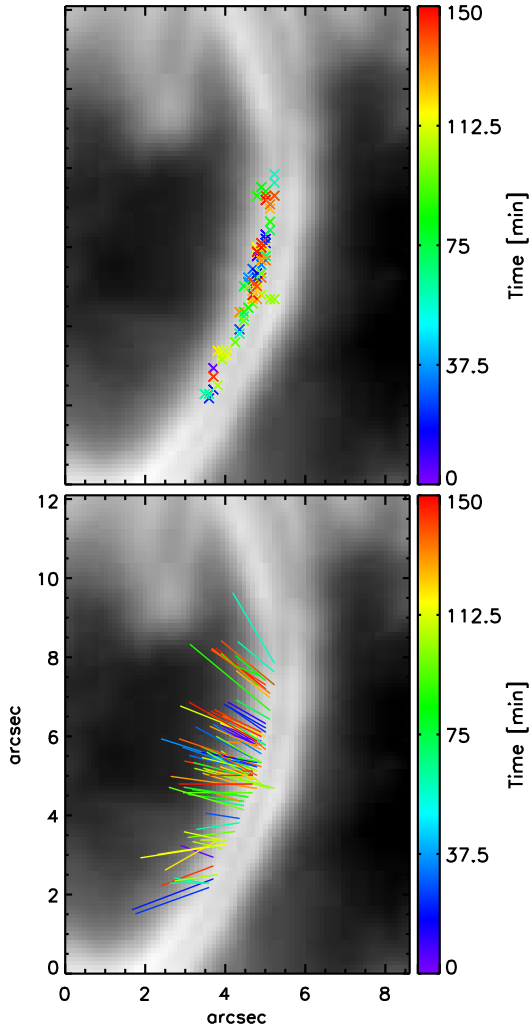}  
}
\vspace{-15pt}
\caption{Representation of the properties of the chromospheric jets detected over the observing period. The 
top panel shows the location of the base of the jet while the bottom panel indicates the length and 
orientation of the jet. The colours represent the time in the sequence when the jet was identified. The 
{\em crosses} and {\em lines} have been overlaid on the time-averaged Ca~{\sc ii}~H image.}
\label{fig03}
\end{figure}

In the above, we assumed that the jet was a linear structure that extends from its base ($x_b,y_b$) into 
the umbra. We set $L$ to 15 pixels and let $\phi_k$ vary from 140$^\circ$ to 220$^\circ$. 
The angle was measured relative to the $x$-axis and was limited to that range to avoid 
including the main body of LB itself because it would have created spurious results. The orientation 
($\phi_k$) of the jet is the angle $\phi_k$ for which $I_k$ is a maximum. Once the orientation was determined 
from the above procedure, the length of the jet was calculated from its base to the end of the jet where 
its intensity fell below a fixed threshold which was essentially the time-averaged umbral background derived 
from the high-pass filtered time sequence. In this manner, three parameters were determined for each jet in 
a given snapshot, i.e. the length, the orientation angle, and position. The jets identified in the sequence 
shown in Fig.~\ref{fig02} are depicted by purple lines that demonstrate the effectiveness of the above algorithm 
as well as the assumption that the extended part of the jets are linear. Since the primary selection of the 
jets was done by hand, there were several frames in the sequence that were skipped because the identification 
of jets was deemed uncertain (e.g. panel 3 in Fig.~\ref{fig02}). We finally determined the orientation 
of the jet with respect to the normal of the LB axis. This axis was derived from the central dark lane that 
runs along the length of the LB that is clearly visible in the time-averaged Ca~{\sc ii}~H image shown in 
Fig.~\ref{fig03}. In the animation it can be seen that the location of the central dark lane was extremely 
stable the whole time with no change at all. We identified 87 jets in 75 Ca~{\sc ii}~H images.

\begin{figure}[!h]
\centerline{
\includegraphics[angle=0,width = \columnwidth]{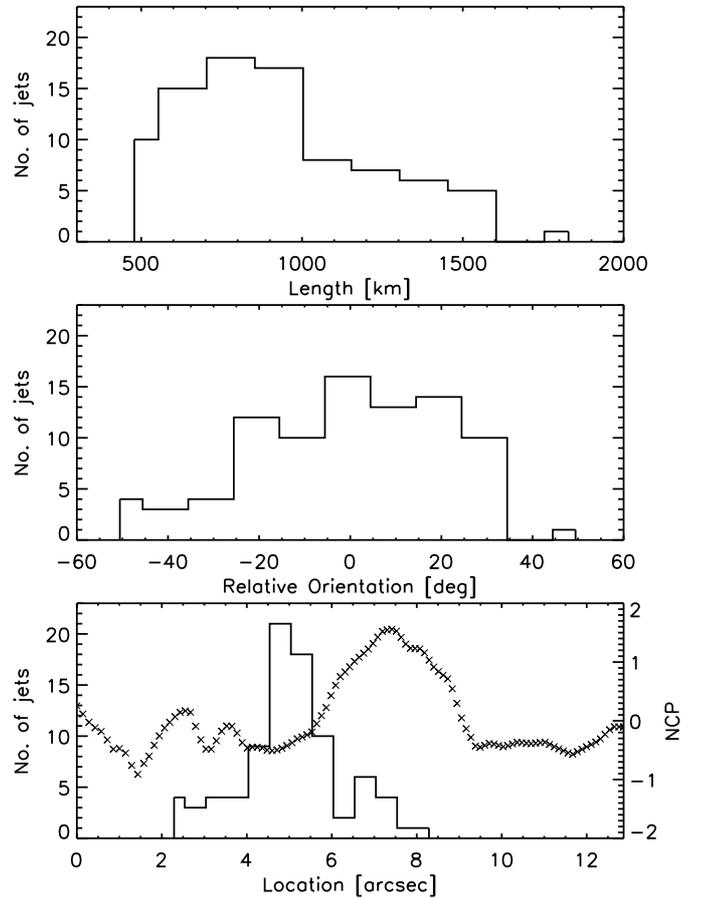}
}
\vspace{-5pt}
\caption{Histogram of jet parameters. Top, middle, and bottom panels correspond to the length, relative orientation, 
and location of the jets, respectively. The bin sizes are 150 km, 10$^\circ$, and 0\farcs5, respectively. In the 
bottom panel the {\em crosses} refer to the net circular polarization along the centre of the LB 
with the scale given at the right-hand side of the plot.}
\label{fig04}
\end{figure}

\begin{figure*}[!ht]
\centerline{
\includegraphics[angle=90,width = \textwidth]{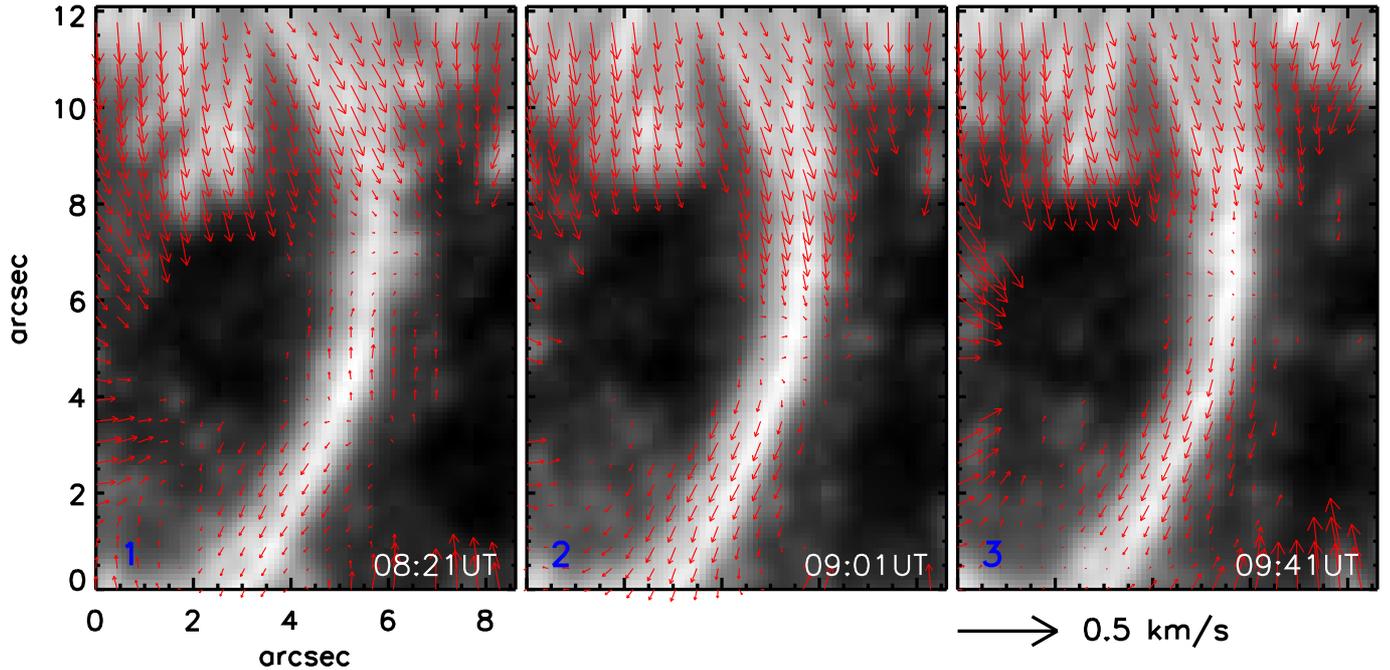}
}
\vspace{-30pt}
\caption{Horizontal flow maps from LCT. The red arrows represent the horizontal flow field that was derived 
from averaging 20 flow maps spanning a duration of 40 min. These have been overlaid on the corresponding time 
averaged G-band image.}
\label{fig05}
\end{figure*}

\section{Results}
\label{result}
\subsection{Properties of jets}
\label{prop}
Although the basic characteristic of the chromospheric jets, described in the previous section and identified 
subsequently, is the spike-like structure with a bright blob on the LB, we could not identify or failed to 
detect any such feature in 20 snapshots spread over the observing run. Panel 3 in Fig.~\ref{fig02} is one 
such example where we find localised blobs of brightness on either side of the central dark lane, but these 
structures do not appear to have the essential property of a jet. We also find that such features without 
a spike exist on both sides of the central dark lane, but the jets we identify all lie on the eastern side, i.e.
on the limb side of the LB. The Ca~{\sc ii}~H movie reveals that the blobs can also appear extended along the length 
of the LB. Some of them seem to move along the LB in the general direction of the photospheric velocity 
field (see Sect.~\ref{lct} below), but there are also cases of motion in the opposite direction.

The top panel of Fig.~\ref{fig03} shows the positions of the bases of the jets on the LB with the colour 
indicating the frame number or time of the snapshot. With the exception of the two anchorage ends of the 
LB, the base of the jets are entirely distributed over its length, although there is a preferential 
clustering near the central part of the LB. The bottom panel of the figure indicates the apparent length 
and orientation of the jets and shows that these features extend from the LB to the eastern umbral core. 

Figure~\ref{fig04} shows the histogram of the apparent length, orientation of the jets relative to the 
normal of the LB axis, and location with bin sizes of 150 km, 10$^\circ$, and 0\farcs5, 
respectively. Most of the jets are shorter than 1000 km with the peak of the distribution lying around 
750 km. Nearly 30\% of the jets detected have an apparent length between 1000--1600 km. The relative 
orientation of the jets, indicated in the bottom panel shows that the jets lie within $\pm$35$^\circ$ 
of the LB axis normal. Here, positive and negative values of the orientation in the figure refer 
to an orientation of the jet to the south and north of the normal to the LB axis, respectively.
The location of the jets is measured along the central axis of the LB with 0\arcsec 
and 12\farcs9 being at the bottom and top of the FOV, respectively. The bottom panel of Fig.~\ref{fig04} 
clearly shows a preferential clustering near the central part of the LB with nearly 65\% of the jets 
located within a 2\arcsec~area. The right $y$-axis refers to the net circular polarization 
along the centre of the LB (cf. Sect.~\ref{magnetic} below).

\subsection{Horizontal flow in the photosphere}
\label{lct}
The time sequence of G-band images shows that there is a patch of brightening that moves from the top 
to the bottom of the LB and lies predominantly on the right half of the LB. Small-scale structures in 
the G-band images also appear preferentially in this section of the light bridge. Faint striations are 
observed perpendicular to the central dark lane of the LB which get pushed along with the material flow. 
The dark lane in the G band is fainter and is shifted more to the left edge of the LB while in the 
Ca~{\sc ii}~H images it is more prominent and centred on the LB. Although the chromospheric jets lack a 
one-to-one association with distinct features in the photosphere, they can sometimes be seen at or close 
to a bright grain on the LB (G-band panels 7 and 9 in Fig.~\ref{fig02}).

Figure~\ref{fig05} shows three horizontal flow maps obtained from LCT, each of which is an average of 
20 maps covering a duration of 40 min. Horizontal motions at the umbra-penumbra boundary have speeds 
of about 200--250 m~s$^{-1}$, while in the LB, they are about 100--150 m~s$^{-1}$. In all three maps, 
the flow is directed into the LB from its northern anchorage end to its southern end. However, the flow 
is neither uniform nor constant along the length of the LB. The flow pattern is interrupted by regions 
with very small or weak motions (see also the animation). The patch of weak flows is again followed by 
the downward flow towards the southern end. This patch of weak flows moves along the LB as it appears 
at the northern end and in the central part of the LB as seen in panels 1 and 2, respectively. The 
horizontal speeds at the southern end are much weaker than at the northern end where the southward flow 
along the LB encounters northward motions from the penumbra into the umbra which is similar to the 
findings of \citet{2008SoPh..252...43L}. The predominantly north-to-south horizontal motion can be 
seen in the G-band movie which also reveals small-scale barb-like features in the LB that evolve 
on timescales of 10--20 min.

\begin{figure}[!h]
\centerline{
\includegraphics[angle=90,width = \columnwidth]{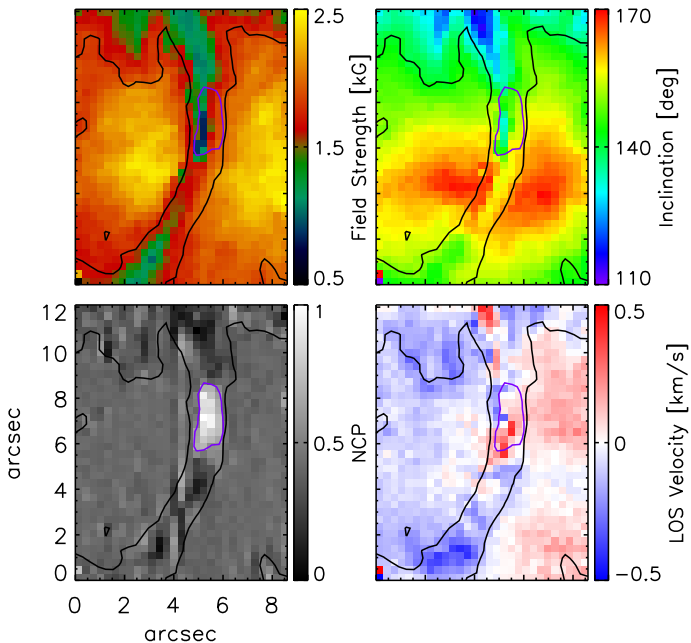}
}
\vspace{-10pt}
\caption{Physical parameters in the light bridge. Clockwise from top left: magnetic field strength, field 
inclination, LOS velocity, net circular polarization. All maps have been scaled according to their colour 
bars. The maps correspond to the SP scan from 10:05--10:21 UT. The purple contour in the panels outlines a 
region of large NCP. The black contour outlines the light bridge and was derived from 
the continuum intensity image at 630 nm in the SP scan.}
\label{fig06}
\end{figure}

\subsection{Magnetic field topology of the LB}
\label{magnetic}
Figure~\ref{fig06} shows the magnetic field strength, inclination, and LOS velocity derived from the inversion 
of the observed Stokes profiles. The figure also shows a map of the net circular polarization (NCP) determined 
by integrating Stokes $V$ over both spectral lines. The field strength in the LB varies along its length (top 
left panel). We find relatively weak magnetic fields of about 1--1.3 kG extending into the LB at its upper and 
lower ends with some pixels showing field strengths of about 800~G  near its upper half. In the central part 
of the LB the field strengths are higher, ranging from 1.5--1.7 kG. We also find a very small region of weaker 
magnetic fields near the central part of the LB close to its left edge bordering the umbra. In the umbral cores, 
where the field orientation is nearly vertical, the field strength is about 2.5~kG. The inclination in the LB 
is about 140--150$^{\circ}$ in the upper end of the LB, while in the central part of the LB the magnetic fields 
are relatively vertical. The location of the weak magnetic fields in the upper half of the LB coincides with 
more inclined magnetic fields, while in the central part the stronger fields are more vertical. The bottom right 
panel of Fig.~\ref{fig06} shows the LOS velocity with the LB exhibiting weak blueshifts but also redshifts 
of about 400~m~s$^{-1}$ at the location of the weak, inclined magnetic fields. At this location we also see 
that the NCP is non-zero within a large patch, which indicates the possible presence of vertical gradients in 
the magnetic field strength, inclination, and/or LOS velocity.

\begin{figure*}[!ht]
\centerline{
\includegraphics[angle=0,width = 0.5\textwidth]{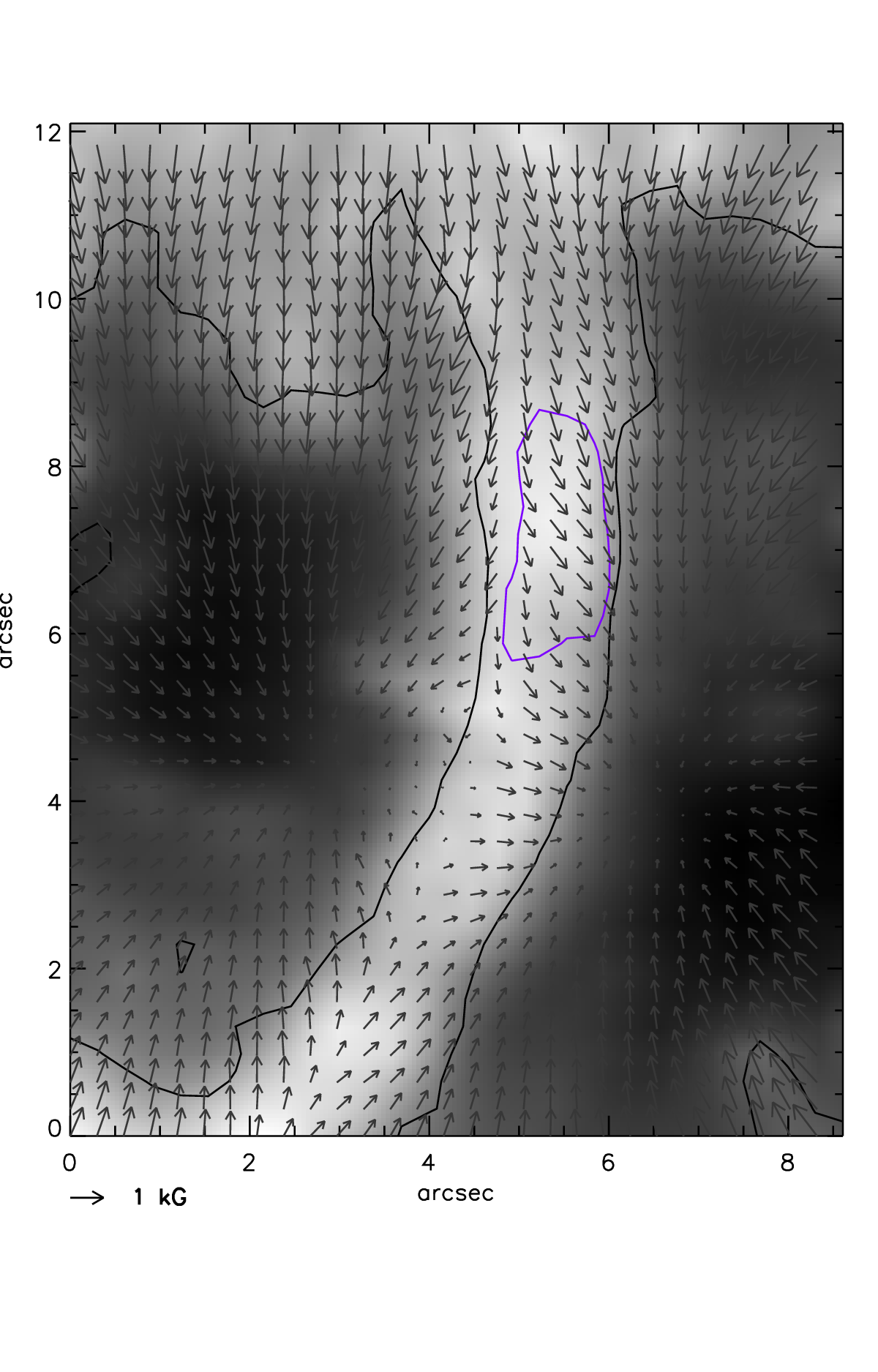}
\includegraphics[angle=0,width = 0.5\textwidth]{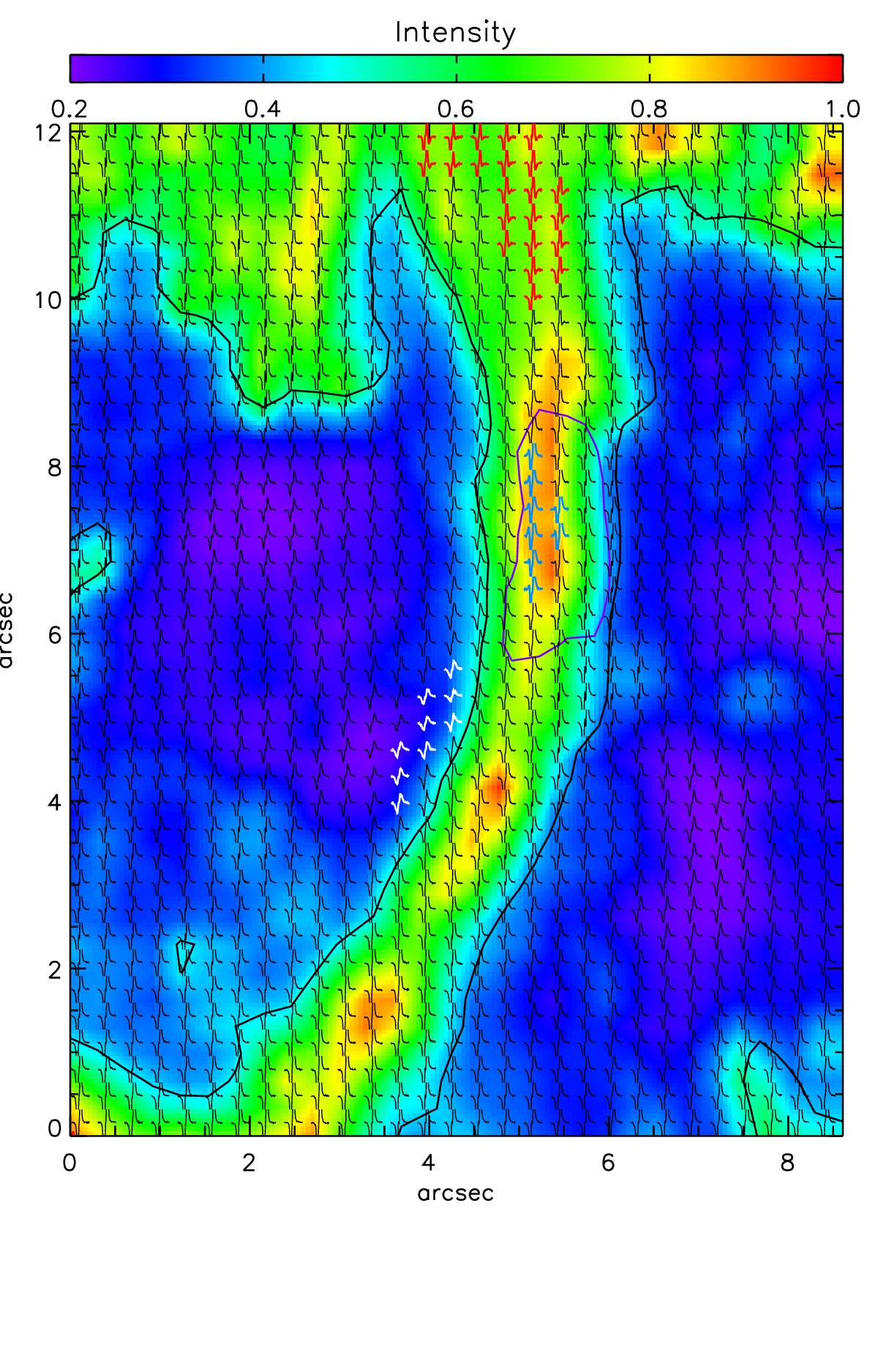}
}
\vspace{-50pt}
\caption{Chromospheric jets and the photospheric magnetic field. The left panel shows the 
horizontal magnetic field from an SP scan with black arrows for every pixel and has 
been overlaid on a Ca~{\sc ii}~H filtergram taken 6 min after the SP slit was above the 
light bridge. The right panel shows the photospheric continuum intensity map in the background 
with the Stokes $V$ profiles overlaid on each pixel of the map. Only the 6302.5 \AA~line has 
been shown and has been clipped to $\pm$8\% of the quiet Sun continuum intensity. 
The purple contour outlines the region of strong NCP indicated 
in Fig.~\ref{fig06}. See text for the description of the blue, red, and white profiles. 
The black contour outlines the light bridge and was derived from 
the continuum intensity image at 630 nm in the SP scan.}
\label{fig07}
\end{figure*}

The left panel of Fig.~\ref{fig07} shows a Ca~{\sc ii}~H filtergram with the horizontal magnetic field overlaid 
on it. It is evident from the figure that the horizontal magnetic field diverges along the edges of the light 
bridge arising from the intrusion of hot plasma in the umbra. The diverging magnetic field at the edge of the 
LB is suggestive of a canopy \citep{2006A&A...453.1079J}. In addition, the magnetic field at the central part 
of the LB is oriented across the LB, while in the lower half of the LB the azimuth changes once again with 
the field directed along the LB. The disruption in the magnetic field also creates two azimuth centres in the 
two respective umbral cores adjacent to the light bridge. The right panel of Fig.~\ref{fig07} shows the 
photospheric continuum intensity map with the Stokes $V$ profile overlaid at every pixel of the image. 
While most of the profiles in the panel are normal profiles with anti-symmetric lobes, there are also 
profiles with a weak third lobe in the blue wing (blue profiles) which are located near the upper half of the 
LB. This additional lobe has a polarity opposite that of the sunspot. We attribute the enhanced NCP (outlined 
by the purple contour) in this part of the LB to the presence of anomalous circular polarization signals. 
The bottom panel of Fig.~\ref{fig04} also indicates that the region of enhanced NCP lies immediately next 
to the location where a vast majority of the jets occur, while a few jets also originate from this patch. This 
suggests that the complex configuration of the magnetic field at this location is responsible for the preferential 
occurrence of the chromospheric jets. At the northern entrance of the LB, there are $V$ profiles that have a similar 
property, but here the third lobe appears on the red wing (red profiles). The blue and red coloured profiles suggest 
that the additional lobe is associated with a magnetic component that is strongly blue- and redshifted, respectively, 
by about 3--4 km~s$^{-1}$. The strength of the lobe also implies that this additional component has a very small fill 
fraction. These pixels also extend out into the northern penumbra. Furthermore, we see $V$ profiles with reduced 
amplitudes (white profiles) that lie exclusively at the left edge of the LB and in the adjacent umbral core. The 
tiny patch of relatively weak fields near the left edge of the LB described above stems from these profiles. We 
also note that the maps shown in Fig.~\ref{fig06} were produced from a Milne-Eddington inversion with a single 
magnetic component. Such an inversion scheme cannot reproduce the spectral characteristics of the blue and red 
coloured profiles shown in Figure~\ref{fig07} where one encounters gradients in the magnetic field and LOS velocity 
or multiple components in the resolution element \citep[see~][and references in the last two articles]
{1978A&A....64...67A,1992ApJ...398..359S,2011A&A...525A.133B,2011LRSP....8....4B}.

\section{Discussion}
\label{discuss}

It is well known that the chromospheres of sunspots exhibit a number of dynamic phenomena. One of 
these is umbral flashes (UFs), sudden brightenings that are observed in the core of the chromospheric 
Ca {\sc ii} lines with a periodicity of about 3 min \citep{1969SoPh....7..366W}. Umbral flashes are 
a result of waves propagating from the photosphere which turn into shocks on reaching the less 
dense chromosphere and are seen as blueshifted emission reversals in the core of the Ca~{\sc ii}~8542~\AA~
line \citep{2003A&A...403..277R,2006ApJ...640.1153C,2010ApJ...722..131F}. 
Although UFs are a large-scale phenomenon, it has been shown that within a resolution element 
there are at least two distinct magnetic components, one that has the same polarity as the spot with 
zero or slightly downflowing velocity, while the other is related to the shock wave that has a line 
core emission reversal and strong upflows \citep{2000Sci...288.1396S}.

The other dynamic phenomenon in sunspots are ubiquitous, fine-scale jets called microjets (MJs) that 
are observed in the chromospheric penumbra \citep{2007Sci...318.1594K}. Microjets have lifetimes of less 
than 1~min and widths of about 400~km. Their lengths vary between 1000--4000 km, although some MJs 
can have lengths of up to 10000~km. It has been suggested that MJs are produced by reconnection in the intermediate 
region between a penumbral flux tube and the relatively vertical, background magnetic field as demonstrated 
in an MHD model by \citet{2010ApJ...715L..40M}. 

In a recent study of six penumbral transients \citet{2013ApJ...779..143R} found that 
some of the events had spectral properties similar to those of Ellerman bombs 
\citep{1917ApJ....46..298E,1956Obs....76..241S,2006ApJ...643.1325F}, with enhanced emission in the 
wings of the Ca~{\sc ii} 8542~\AA~line. Other transients identified were possibly related to strong 
acoustic shocks, umbral flashes, or an emerging magnetic bipole in the chromosphere. Such emission 
features in the different parts of the spectral line show up as brightness enhancements in the broadband 
{\em Hinode} Ca~{\sc ii}~H filtergrams. From their observations, \citet{2013ApJ...779..143R} 
suggest that the complexity and variety of the magnetic field structure in different regions of the 
sunspot is not only related to the distinct spectral characteristics of the transients, but also to the rapid 
timescales on which magnetic energy is released in the solar atmosphere.

The small-scale jets we have described could be produced by reconnection at the edge of the light bridge where 
the magnetic field is seen to diverge or wrap around it. The magnetic field also changes orientation along the 
length of the LB where the magnetic fields from the north encounter oppositely directed fields from the south 
near the central half of the light bridge, similar to the observations of \citet{2009ApJ...704L..29L}. 
Anomalous Stokes $V$ profiles having multiple components and lobes of opposite polarity are 
another indicator that the magnetic configuration in the LB is complex \citep[see also][]{2007MNRAS.376.1291B}.
These profiles (coloured blue at the middle of the LB in Fig.~\ref{fig07}) indicate the presence of strong 
gradients in the magnetic field and LOS velocity and/or multiple components in the resolution element. This 
location of enhanced net circular polarization lies in the immediate vicinity of where most of the jets 
occur, and strongly suggests that the stressed configuration of the magnetic field is responsible for the 
preferential occurrence of the chromospheric jets. 
\citet{2009ApJ...696L..66S} showed that the presence of strong electric currents exclusively 
on one edge of a light bridge was responsible for chromospheric ejections being launched dominantly from 
that side of the light bridge. They interpreted the light bridge as a current-carrying twisted flux tube 
trapped below a cusp-shaped magnetic structure where strong currents were produced at the interface of 
the tube from fields having an antiparallel orientation with respect to the neighbouring umbral field. 

In addition to the arrangement of the magnetic field, we also find a unidirectional, non-uniform 
flow that could facilitate the nearly persistent nature of the jets. Since LBs are a manifestation of weakly 
magnetised plasma disrupting the pre-existing nearly, homogeneous umbral magnetic field, the extent of the 
chromospheric activity could be related to the extent of the perturbation of the umbral magnetic field. This is 
demonstrated in individual case studies of LBs that at least show some related chromospheric activity at a 
certain stage of their evolution. On the other hand, the 
unidirectional, non-uniform flow in the light bridge could perturb the magnetic field in the light 
bridge exciting waves that propagate to the higher solar atmosphere, releasing the energy in the 
form of the jets. The coarse 
temporal sampling of the data and the limitations of our detection scheme mean that some of the jets are excluded 
from the study which could influence even static properties, namely the position, length, and relative orientation.
Since we do not have the complete spectral line information, it is not possible 
to ascertain if the jets are associated with a line-core emission or an enhancement in the overall 
spectral intensity. We also note that a 0.3 nm filter is used for broadband Ca~{\sc ii}~H imaging, as is the case for 
{\em Hinode}, and primarily samples the higher photospheric layers less than 600~km in height, 
which is dominated by reverse granulation and not from higher chromospheric layers above 1000~km, for 
observations at disc centre \citep{2013A&A...556A.127B}. Thus, even for our observations at a 
heliocentric angle of 30$^{\circ}$, it is possible that the jets we have described are more 
upper photospheric in nature than fully chromospheric. Instruments such as the Blue 
Imaging Solar Spectrometer \citep[BLISS,][]{2013OptEn..52h1606P} that has been planned at the 
1.5 m GREGOR telescope \citep{2012AN....333..796S}, the Interferometric BIdimensional 
Spectrometer \citep[IBIS;][]{2006SoPh..236..415C}, and the CRisp Imaging SpectroPolarimeter
\citep[CRISP;][]{2008ApJ...689L..69S} will be crucial in understanding the mechanism responsible 
for these small-scale jets.

\section{Conclusions}
\label{conclu}
The small-scale, short-lived, chromospheric jets above the sunspot light bridge seem to be guided by the 
magnetic field. Reconnection events are a likely trigger for these phenomena since they occur at locations where the
magnetic field changes orientation sharply. Localized patches of opposite-polarity magnetic components further
illustrate that the magnetic field in the light bridge is complex which lends support to the reconnection scenario. 
While we do not find a clear relation between the chromospheric jets and the photospheric flow pattern, we do not rule
out the possibility of the latter's role in perturbing the magnetic field and subsequently producing the jets through
magnetic reconnection or by the excitation of waves. On the observational side, high-resolution spectroscopy in the 
chromosphere with high temporal resolution is crucial for determining the thermal and kinematic properties of 
such jets, which is not possible with our current broadband imaging observations.

\begin{acknowledgements}
Hinode is a Japanese mission developed and launched by ISAS/JAXA, collaborating with NAOJ as a 
domestic partner and NASA and STFC (UK) as international partners. Scientific operation of the 
Hinode mission is conducted by the Hinode science team organized at ISAS/JAXA. This team mainly 
consists of scientists from institutes in the partner countries. Support for the postlaunch 
operation is provided by JAXA and NAOJ (Japan), STFC (UK), NASA, ESA, and NSC (Norway). 
Hinode SOT/SP Inversions were conducted at NCAR under the framework of the Community 
Spectro-polarimtetric Analysis Center (CSAC; http://www.csac.hao.ucar.edu/). R.E.L is 
grateful for the financial assistance from the German Science Foundation (DFG) under grant 
DE 787/3-1. We thank the anonymous referee for the useful comments.
\end{acknowledgements}

\end{document}